\documentclass{article}
\usepackage{nips13submit_e,times}
\usepackage[german,american]{babel}
\usepackage[a4paper,top=3.5cm,bottom=3.5cm,left=3.0cm,right=3.0cm,marginparwidth=1.75cm]{geometry}
\usepackage{amsmath}
\usepackage{amsfonts}
\usepackage{amssymb}
\usepackage{graphicx}
\usepackage[font=small,labelfont=bf]{caption}
\usepackage{subcaption}
\usepackage{float}
\usepackage[OMLmathsfit]{isomath}
\usepackage{amssymb} 
\usepackage{amsxtra} 
\usepackage{bm} 
\usepackage{mathtools}
\usepackage{verbatim}
\usepackage[colorlinks=true, allcolors=blue]{hyperref}
\usepackage{lineno}
\usepackage{soul}
\usepackage{xcolor}

\nipsfinalcopy % Uncomment for camera-ready version
\DeclareCaptionLabelFormat{mylabel}{#1 #2\hspace{1.5ex}}
\usepackage[numbers,sort&compress]{natbib}

\definecolor{darkgreen}{rgb}{0.0,0.5,0.0}
\definecolor{BurntOrange}{rgb}{0.8,0.3,0.0}
\definecolor{mygray}{gray}{0.5}

\usepackage{booktabs}

\newcommand{\figref}[1]{Fig.~\ref{#1}}
\newcommand{\equref}[1]{Eq.~(\ref{#1})}
\newcommand{\secref}[1]{Sec.~(\ref{#1})}

\providecommand{\keywords}[1]
{
  {
  \small	
  \textbf{\textit{Keywords---}} #1
}}

\title{Computational Discovery of Energy-Efficient Heat Treatment for Microstructure Design using Deep Reinforcement Learning}

\author{Jaber R. Mianroodi$^{1,*}$, 
Nima H. Siboni$^{1}$, Dierk Raabe$^{1}$\\
       \footnotesize $^1$Microstructure Physics and Alloy Design, \\ 
        \footnotesize Max-Planck-Institut f\"ur Eisenforschung, D\"usseldorf, Germany \\
        \footnotesize $^*$Corresponding Author j.mianroodi@mpie.de
} 

\date{\today}

\begin{document}

\maketitle

\begin{abstract} 
Deep Reinforcement Learning (DRL) is employed 
to develop autonomously optimized and custom-designed heat-treatment processes that are both, microstructure-sensitive and energy efficient.
Different from conventional supervised machine learning, DRL does not rely on static neural network training from data alone, but a learning agent autonomously develops optimal solutions, based on reward and penalty elements, with reduced or no supervision. This approach resembles human learning more than other methods used in artificial intelligence.
In our approach, a temperature-dependent Allen--Cahn model for phase transformation is used as the environment for the DRL agent, serving as the model world in which it gains experience and takes autonomous decisions. The agent of the DRL algorithm is controlling the temperature of the system, as a model furnace for heat-treatment of alloys. Microstructure goals are defined for the agent based on the desired microstructure of the phases. After training, the agent can generate temperature-time profiles for a variety of initial microstructure states to reach the final desired microstructure state. The agent’s performance and the physical meaning of the heat-treatment profiles generated are investigated in detail. In particular, the agent is capable of controlling the temperature to reach the desired microstructure starting from a variety of initial conditions. This capability of the agent in handling a variety of conditions paves the way for using 
such an approach also for recycling-oriented heat treatment process design where the initial composition can vary from batch to batch, due to impurity intrusion, and also for the design of energy-efficient heat treatments. For testing this hypothesis, an agent without penalty on the total consumed energy is compared with one that considers energy costs. The energy cost penalty is imposed as an additional criterion on the agent for finding the optimal temperature-time profile. 

\end{abstract}


\section{Introduction}
Macroscopic material properties, in particular the mechanical properties of metallic alloys, are controlled by the distribution of small-scale crystalline defect features, known as microstructure, and their interactions. Therefore, it is necessary to design the underlying microstructure to achieve desired alloy properties. This is typically achieved by varying the chemical composition (thus enabling and tuning different solid solution states, kinetic parameters, and phases), thermal, and/or mechanical processes. Discovering new compositions or thermo-mechanical processes to reach desired microstructures and properties is one of the most complex problems in material science. Historically, these discoveries were driven by luck, serendipity, or experimental trial and error. 

Recently, computational methods have been employed to accelerate material and process discovery. Machine learning (ML) as a tool to accelerate material modeling has proven to be an effective method in that context \cite[e.g.][]{ali2019application,mayer2021dislocation,pandya2020strain,settgast2020hybrid,mianroodi_lossless_2022,kobeissi_enhancing_2022,prachaseree_learning_2022,mohammadzadeh_predicting_2022}. ML-based methods have been also employed extensively for metallography, pattern recognition, localization detection, microstructural characterization and damage analysis \cite{mangal_applied_2019,mangal_applied_2018,cohn_unsupervised_2021,dimiduk_perspectives_2018,holm_overview_2020}. 
Most of the works on the application of ML in material modeling can be categorized into the development of interatomic potentials \cite[e.g.][]{MISHIN2021116980}, homogenization of material response \cite[e.g.][]{settgast2020hybrid,mianroodi_lossless_2022}, and surrogating material models and partial differential equations \cite[e.g.][]{mianroodi2021teaching,Khorrami2022}. Almost all of these works are focused on material modeling, design, and the relationship between microstructure and property. Although material processing is as important as the composition in determining the property of the alloy, fewer works have been carried out on using ML for the computational design of the processing pathways.
As a recent example, Dornheim et al. \cite{dornheim_deep_2022} used reinforcement learning (RL) for path optimization of metal forming processes. A few other examples of the applications of RL in material science are also available in the literature for designing bio-inspired structural composites \cite{yu_hierarchical_2022} and magnetic control of plasma \cite{degrave2022magnetic}. 

Heat treatment of alloys is one of the main processing steps to control, design, and even repair the microstructure, that is inherited from the preceding processing steps, particularly casting, homogenization and plastic deformation.
In addition heat treatment is also one of the most energy-intensive processing steps in the metallurgical sector, accounting for 1 to 2 percent of the total global energy consumption. In the past, heat treatments were mostly developed to meet certain microstructure-oriented needs alone, yet, spiking energy costs and the quest for a more sustainable industry create a high motivation and leverage to optimize heat treatments for both, microstructure and properties and also for reduced energy consumption. Often alloys with the same composition exhibit properties with orders of magnitude variations simply due to differences in heating and cooling process steps including both different temperatures, holding times, and rates. In terms of recyclability and circular economy, designing material properties through heat treatment optimization is more beneficial compared to increasing the compositional complexity. However, most of the heat treatment curves have rather simple profiles, developed essentially by experimental methods and - in part - by mean-field kinetic and thermodynamic simulations. A more thorough, energy efficient, and microstructure-oriented search of the possible temperature profiles for discovering novel microstructures has been less in the focus so far. 

In this work, we present a showcase of employing deep RL for discovering heat treatment. We couple a temperature-dependent phase-field (PF) model, designated as the learning environment for the RL approach, with a neural network-based agent. The agent learns to decide which actions to select based on the microstructure state and the goals defined. The actions available for the agent include increasing or reducing the temperature, keeping the temperature constant, or ending the heat treatment. Different goals based on the microstructure as well as the energy consumption of the heat treatment process are included in the unsupervised learning. The trained agent is then evaluated on a set of test data, with different variations in initial conditions. In Section \ref{sec_pf}, details of the PF model, which acts as the environment for the agent in our RL approach, are summarized. This is followed by the description of the RL algorithm in Section \ref{sec_rl}. Unsupervised training of the agent is explained in Section \ref{sec_training}. Results and benchmarking of the trained model are discussed in Section \ref{sec_res} and the paper is closed by a conclusion in Section \ref{sec_conc}. 

\section{Phase-field method}
\label{sec_pf}

The phase-field method is one of the most efficient, versatile, and physics-wise well-founded approaches for investigating moving boundary problems, which qualifies it as a suited approach for microstructure evolution simulations. Several examples of the application of PF to material modeling are available in the literature \cite{khorrami_finite2022,bai_chemomechanical_2022,mianroodi_phasefield_2021}. Here for simplicity and proof of concept, we employ a simple Allen--Cahn PF model
\begin{equation}
\dot\phi = -M \frac{\delta \psi}{\delta \phi},
\label{eqAC}
\end{equation}
where $\phi$, $M$, and $\psi$ are the order parameter, mobility, and free energy density, respectively. The order parameter is a scalar field representing the microstructure. In our case, $\phi=0$ represents the first phase while $\phi=1$ is the second one. The total free energy of the system 
\begin{equation}
\Psi(\phi,\nabla \phi, T) = \int_{\Omega} \psi(\phi,\nabla \phi, T) d\Omega
\label{eqEn}
\end{equation}
is the volume integral of the free energy density ($\psi$) and assumed to be temperature ($T$) dependent in this work. The free energy density consists of the homogeneous part that represents the energy of the phases and the gradient part that represent the interfacial energy contributions
\begin{equation}
\psi(\phi,\nabla \phi, T) = \psi_\mathrm{H}(\phi,T) + \psi_\mathrm{gr}(\nabla \phi).
\label{eqEnDiv}
\end{equation}
Substituting the equation above in \eqref{eqAC} and \eqref{eqEn} and carrying out the variational derivative we get: 
\begin{equation}
\dot\phi = -M (\frac{\partial \psi_\mathrm{H}}{\partial \phi} - 2\kappa \nabla^2 \phi),
\end{equation}
for the time evolution of the order parameter. The forms of the homogeneous and the gradient energy contributions used in this work are 
\begin{equation}
\begin{array}{ccl}
 \psi_\mathrm{H}(\phi, T) &  =  & a_4(T) \phi^4 + a_3(T) \phi^3 + a_3 \phi^2 + a_1(T) \phi + a_0(T) \\
  \psi_\mathrm{gr}(\nabla \phi) & =   & \kappa |\nabla \phi|^2
\end{array}
\label{eqhomgrad}
\end{equation}
with coefficients $a_i$ and $\kappa$ listed in Table \ref{tab1}. The polynomial approximation for the homogeneous energy part follows the classical Landau energy ansatz with one stable phase at a value of zero for the order parameter and the other phase being stable when the order parameter assumes a value of 1, and the balance depends on temperature.
\begin{table}[htb]
    \centering
    \caption{Coefficients of the free energy density as well as the mobility model used in the phase-field simulations. $a_\mathrm{i}$ are in units of $\mathrm{J/m^3}$.}
    \begin{tabular}{|c|c|c|c|c|c|c|c|}
    \hline
           $a_4 $ & $a_3$ & $a_2 $ & $a_1$ & $a_0 $ & $\kappa\, (\mathrm{J/m})$ & $A\,(\mathrm{m^3/J^{-1}s^{-1}})$  & $Q\,(\mathrm{meV})$  \\ \hline
         0.006T+47   &  -0.008T-96     & 50.0  &   0.0  &   0.001T     & 30.0 & 2.8$\times 10^{-4}$ & 2.9\\ \hline
    \end{tabular}
    \label{tab1}
\end{table}

\begin{figure}[htb]
    \centering
    \includegraphics[width=0.49\textwidth]{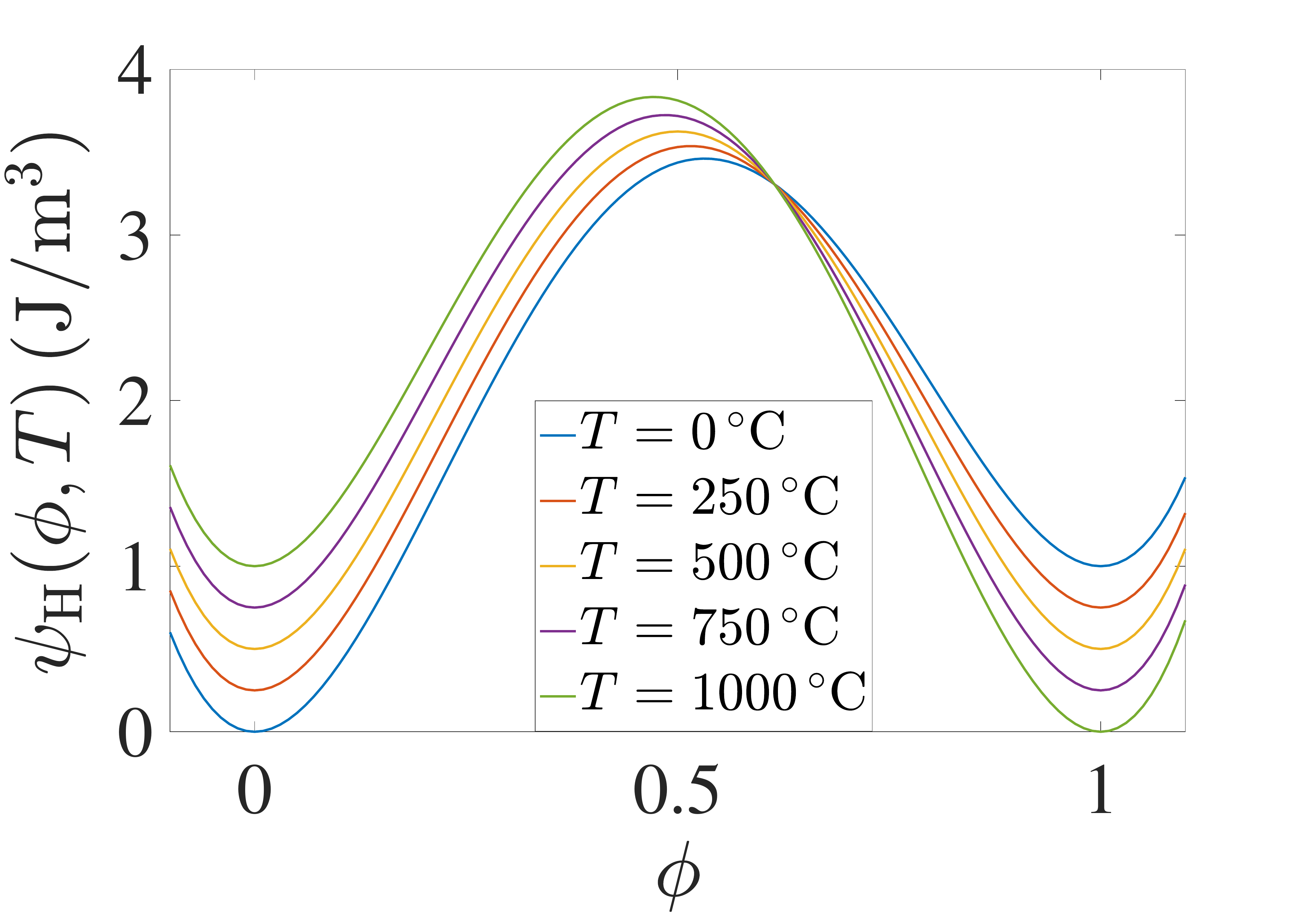}
    \includegraphics[width=0.49\textwidth]{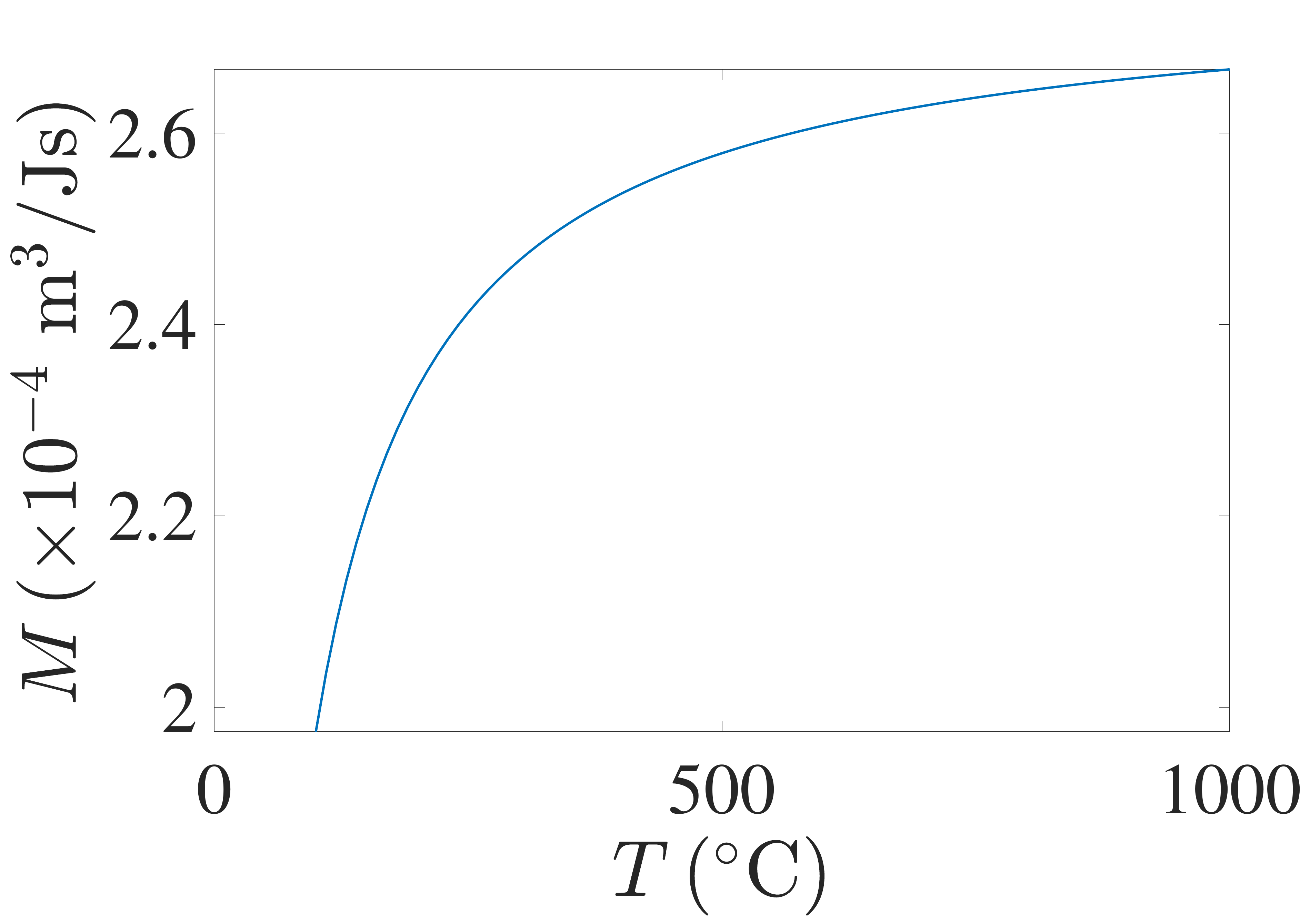}
    \caption{Left: Homogeneous part of the free energy density $\psi_\mathrm{H}(\phi, T)$ (see \eqref{eqhomgrad}) at different temperatures. Right: Interface mobility as a function of temperature.}
    \label{figbulkmob}
\end{figure}

Employing the coefficients listed in Table \ref{tab1}, the homogeneous part of the free energy density ($\psi_\mathrm{H}$) is plotted for several temperature levels in Figure \ref{figbulkmob}. As it is seen in this figure, at low temperatures, phase 1 ($\phi=0$) has lower energy and is more stable, while at high temperatures, phase 2 ($\phi=1$) is the stable phase. At $T=500^\circ$C the two phases have the same energy. 

The mobility $M$ in \eqref{eqAC} is assumed to be temperature dependent, following the classical Turnbull rate theory which balances the forward and backward jumps of the interface-adjacent atoms through the interface, penalized by a barrier $Q$ according to
\begin{equation}
   M = A \exp{(\frac{-Q}{k_\mathrm{B} T})}
   \label{eqMob}
\end{equation}
with coefficients $A$ and $Q$ listed in Table \ref{tab1} and $k_\mathrm{B}$ representing the Boltzmann constant. The interface mobility as a function of temperature is plotted in \figref{figbulkmob} on the right-hand side. As in most transport and phase transformation problems of this kind, the interface mobility increases with increasing temperature, and the actual net interface motion results from the driving force that acts on it. In the presented form of the total energy functional, this driving force comes from the temperature-dependent difference in the Landau energy and the reduction of the interface-related energy (gradient) term, i.e. the reduction of the local interface curvature.

\section{Reinforcement learning framework}
\label{sec_rl}
\subsection{Background}
Mechanical and thermal heat treatments in material science are examples of sequential decision-making processes. In sequential decision-making, as the name suggests, an objective is achieved by collecting information and taking a sequence of actions. The system starts in a state, and in each "step" one takes a specific decision and action which consequently changes the state of the system to a new state from which the next step is taken. Additional information is gathered along the sequence of steps and considered in the ensuing decision-making and action steps. Such sequences of information-decision-making and action processes are ubiquitous, and not limited to material engineering processes. An illustrative example is navigation on a map where the objective is to arrive at the destination as quickly as possible. In this case, the state is our current position on the map (i.e. at which junction we are) and the actions are which direction to go (e.g. taking right, left, forward, or backward turns). By taking an action, we travel to the next junction, where the duration of this travel depends on the current state and the action we chose. One can express the objective of the problem, i.e. getting as quickly as possible to the destination, as a minimization of the cumulative sum of the travel times after each action. Such expression of the objective, when possible, becomes useful in finding the optimal behavior in sequential decision-making processes, as explained in more detail below.

A simple, yet general framework for the formulation of sequential decision-making problems is the Markov decision process (MDP). In principle MDP formalizes the example that has been given above, yet, with two additional features that enter the decision-making sequence,  namely,  (a) that the transition between states,  based on taking a specific action, is associated with a certain probability function and (b)  that actions have consequences, in terms of a reward or cost feature. When rendering this formal, MDP is presented by the tuple $\mathcal{M} = \{ \mathcal{S}, \mathcal{A}, P, R\}$, where
\begin{itemize}
    \item $\mathcal{S}$ is the state space, e.g. the set of all the junctions in the navigation problem,
    \item $\mathcal{A}$ is the action space, e.g. set of all the directions available to take from each junction,
    \item $P=P_a(s, s')$ is the Markov transition probability distribution, i.e. the probability of reaching state $s'$ from state $s$ taking action $a$,
    \item $R=r(s, a)$ is the reward/cost associated with taking action $a$ from state $s$, e.g. the travel time from a junction to another junction by taking a particular turn.
\end{itemize}

The goal, when the \textit{reward hypothesis} applies \cite{sutton1998introduction}, is to maximize the cumulative sum of all the rewards taken from the initial state to the last state in the sequence of taken steps.
Among the wide range of possible approaches to find the optimal behavior in an MDP, i.e. the behavior which results in maximization, we choose deep reinforcement learning (DRL) in this work. 

\subsection{Deep reinforcement learning} 
\label{drl}
In reinforcement learning, two separate entities are involved in the sequential decision-making process: (i) the \textit{agent} which is the decision maker (and also the entity which learns the optimal behavior), and (ii) the \textit{environment}, which is everything outside of the agent with which the agent interacts for accumulating information, i.e. creating an experience that is specific for this environment. In the following, some key elements of a DRL setup are briefly outlined and the notion of 'optimality' is defined. 

A central aspect of the decision-making process is the so-called 'policy'. This term defines the learning agent's response when in a given state, i.e. a policy is an algorithm of how perceived states of the agent's environment are translated into actions to be taken when in those states. This means that the agent holds and follows the \textit{policy}, $\pi$, which is a mapping from states to actions. In other words, the agent uses its policy to decide which next action it should take from each state. The policy can be deterministic, $a=\pi(s)$ or it can be probabilistic, i.e. it gives the probability distribution of choosing action $a$ from state $s$, $\pi(a|s)$. Commonly, in a reinforcement learning solution, the agent starts with a random policy and improves the policy using interactions with the environment, as explained below. 

Having a policy, the agent can gather experience via interaction with the environment. In particular, starting from state $s$, the agent takes an action $a$ based on its policy $\pi$ and passes that action to the environment. In return, the environment informs the agent about (i) the \textit{immediate reward} $r(s, a)$ and (ii) the new state in which the agent is after its last action. This process continues until the agent reaches a \textit{terminal condition} where the process ends. The sequence of states from the initial state to the terminal state is referred to as an \textit{episode}. Here, we limit ourselves to tasks that have a terminal condition, and hence episodes are finite.

In DRL, the goal is to find the policy that is optimal for a given environment. As mentioned above, the agent commonly starts with a random policy, gathers experience via its interaction with the environment, and improves its policy until it converges to the optimal policy (or gets close enough). To define the optimal policy, we first need to define the \textit{value} of each state $s$ under policy $\pi$, $V_\pi(s)$. The value of a state under policy $\pi$ is the cumulative sum of the rewards it gets when following that particular policy starting from that state. For a deterministic policy, this reads as
\begin{align}
    V_\pi(s_0) = \sum_{t=0} r(s_t, a_t)~,
\end{align}
where subscript $t$ denotes the $t$-th step, $a_t=\pi(s_t)$ is the action taken from state $s_t$, and the sum runs until the end of the episode. Commonly, the terms in the summation are weighted by \textit{discounting} factors, which reduces the weight of future rewards. The discounting is commonly implemented by a geometrically decreasing factor,
\begin{align}
    V_\pi(s_0) = \sum_{t=0}\gamma^t r(s_t, a_t)~,
\end{align}
where $0\leq \gamma \leq 1$ is the \textit{discount factor}.

Using the value function, we can proceed to define the optimal policy. Policy $\pi$ is \textit{better} than (or equal to) policy $\pi'$, if the value of each state under policy $\pi$ is larger than (or equal to) the value of that state under policy $\pi'$, i.e. $\pi \geq \pi'$ if
\begin{align}
\forall s \in \mathcal{S}: V_{\pi}(s) \geq V_{\pi'}(s)~.
\end{align}
and $\pi^*$ is the optimal policy if 
\begin{align}
\forall \pi': {\pi^*}(s) \geq \pi'~.
\end{align}

Equipped with this definition, one can find out if a modification to the current policy has led to an improvement (i.e. a better policy). There are numerous algorithms for improving the agent's policy from the experience gathered during the application of a non-optimal policy. For a (non-exhaustive) list of commonly used algorithms, we refer the reader to the RLlib's implemented algorithms \cite{rllibalgo}. The choice of the algorithm depends on many factors including the structure of the action space (e.g. whether the action is continuous or discrete), the dimensionality of the state space, the availability and the efficiency of a simulator for building the environment, etc. Our choice for the learning algorithm, in this work, is the Deep Quality Network (DQN), often also referred to as Deep Q-Network \citep{mnih2013playing,mnih2015human}. Since its success in achieving human-level expertise in playing Atari games  \cite{mnih2015human}, DQN and its variations have been one of the most commonly used DRL algorithms (See Ref.~
\cite{li2017deep} and the references therein). In the following, we briefly sketch the basics of the DQN algorithm. 
Note that for the sake of simplicity of notation and argument some simplifications are considered, including restrictions to deterministic policies. For a more detailed introduction, we refer the readers to Ref.~\cite{sutton1998introduction}.

\subsection{The Deep Quality Network (DQN) algorithm}
In the framework of the DQN algorithm, or in general any Q-learning algorithm, instead of directly finding the optimal policy, we find the Q-values of the optimal policy. The Q-value of $(s, a)$ for any policy $\pi$ is the sum of two terms:
\begin{itemize}
    \item the immediate reward after taking action $a$ from state $s$, i.e. $r(s, a)$, and
    \item the cumulative sum of the rewards obtained from the new state when the policy $\pi$ is followed, which is equal to $V_{\pi'}(s')$ with $s'$ being the new state after action $a$ from state $s$.
\end{itemize}
One can note that there is a close relation between $Q_\pi(s, a)$ and $V_\pi(s)$. $Q_\pi(s, a)$ is the total sum of all the rewards from state $s$ when the first action is $a$ and then the policy $\pi$ is followed for the later actions. Only if the first action was from the policy, the value of $Q$ would be identical to $V_\pi(s)$. Note that $Q$ values for different actions from each state reflect the \textit{quality} of those actions from that particular state: the larger the $Q_\pi(s, a)$ the better the action, i.e. that action leads to a larger total reward when followed by the actions from policy $\pi$. Knowing the $Q$ values for a policy offers valuable information on how to \textit{improve} upon that policy, as explained below. This means that a larger Q value quantifies how much more useful a given action is in gaining also future rewards, i.e. it is a quantitative and predictive measure of action quality.

The policy improvement, also known as \textit{policy control}, can be a straightforward procedure when all the $Q$-values are known. For example, if one changes the policy $\pi$ in one arbitrary point (let's call it $s_0$) to $\mathrm{argmax}_a Q_\pi(s_0, a)$, it is straightforward to show that the resulting policy is better than the policy $\pi$ (refer to the definition of a better policy in \ref{drl}). If one repeats this procedure of policy improvement, under certain conditions which are discussed in detail in Ref.~\cite{sutton1998introduction}, one converges to the optimal policy.
 
A prerequisite for a policy improvement step, as sketched above, is to know the $Q$-values of the policy. The procedure of finding $Q$-values of a policy is referred to as \textit{policy evaluation}. There is a variety of policy evaluation methods in which the values are learned from experience. These methods differ from each other in aspects like sample efficiency, induced bias, variance, etc. For a review of these methods refer to Ref.~ \cite{dann2014policy}. The Temporal Difference (TD) method is one specific family of such evaluation methods. In TD, one updates the estimated values of all states iteratively and incrementally. This is, in its essence, similar to a fixed-point iterative scheme where the estimate of the unknown values is updated at each iteration based on the estimated values of the same unknown from the previous iteration. In the case of TD, more specifically, we improve the estimate of the value of the current step based on the estimated values of the next step in the episode, i.e. one \textit{bootstraps} (refer to  SARSA~\cite{sutton1988learning} as a well-known example of such a method). In Q-learning, the evaluation and improvement steps are combined into one update rule \citep{watkins1992q}:
\begin{equation}
Q (s,a) \leftarrow Q(s, a) + \alpha \bigl[ r(s, a) + \gamma  \max_{a' \in \mathcal{A}} Q (s',a') - Q(s, a)\bigr],
\end{equation}\label{eq:Q}
where $\alpha > 0$ is the learning rate. Under certain conditions, e.g. with a finite chance of visiting every state-action pair and the feasibility of independent updates of the Q-values for each $(s, a)$, one can show that the update rule given above converges to the optimal Q-values, i.e. to those Q-values that constitute the optimal policy~\cite{watkins1992q}. Note that the latter condition is easily satisfied if the Q-values are restored in a table, i.e. tabular Q-learning. Nevertheless, in the case of learning scenarios associated with a continuous state space, or discrete state space with a large number of states, using a tabular representation of Q-values is not feasible, and one needs to use a function in such cases. DQN, the method of our choice in this work, is one of the algorithms where Q-values are not presented as a table.

DQN \citep{mnih2013playing,mnih2015human} is a specific variant of Q-learning where the $Q(s, a)$ is represented by a function which is a deep neural network. Using deep neural networks for representing Q not only removes the restrictions of the tabular Q-learning but also utilizes the approximation and generalization power of deep neural networks in the context of reinforcement learning (see discussions in Ref.~\cite{li2022deep} for more insight). Using deep neural networks, from a technical point of view, has another advantage as well: the update in \equref{eq:Q} can be reformulated into a supervised deep learning problem, and therefore all the existing software frameworks and the dedicated hardware of deep supervised learning can be utilized for deep reinforcement learning.

Although using deep neural networks plays an important role in the success of DQN, there are subtleties involved which are the focus of current research in DRL. The origin of these subtleties can be traced back (partially) to the following issues: (i) the condition of independent updates is not anymore satisfied when $Q$ is presented by a function, and (ii) when the update of $Q$-values is cast into a supervised learning problem, the target values (or true values) used for the update of the neural network parameters becomes \textit{dependent} on the $Q$ network itself, i.e. we have a moving target.  These subtleties underpin the convergence guarantees existing for the tabular $Q$-learning update and also make the training process unstable. To remedy these conditions, many different extensions of the DQN method have been introduced (for examples see Refs. ~\citep{van2010double,  van2016deep, schaul2015prioritized, wang2016dueling, sutton1998introduction, bellemare2017distributional, fortunato2017noisy} and references therein). In this work, we use a combination of the aforementioned extensions, which is commonly referred to as the "rainbow"~\citep{hessel2018rainbow} algorithm.

\subsection{Environment and reinforcement learning formulation}
\label{env}
The dynamics of the environment develop in our case following the phase-field model explained in \secref{sec_pf}. For the design of the state space, action space, and rewards, we made choices, whose advantages and disadvantages are discussed in this section. A python implementation of this environment, named "Furnace", can be found in the GitHub repository \cite{SIBONI_Furnace_2022}.

For the state space, instead of a characterization of the micro-state, we directly use the raw high-resolution image of the microstructure itself.  Characterization, roughly speaking, refers to measuring statistical properties of the microstructure or the macroscopic behavior, from which one can infer information about the microstructure, for instance via inverting established microstructure-property relations. In material science and engineering, using characterization for the quantification of microstructure features (such as grain size, dislocation density, etc.) is more common than using microscopic images directly. Here, we choose to follow the latter option, i.e. we use high-resolution microscopic information for the following reasons:
Firstly, there might be information that the agent could potentially learn from the raw image which might not be directly revealed to the same level of information content from a set of limited statistical characterization features otherwise used to describe the state. In other words, characterization (which can be considered feature-engineering in a machine learning context) has the potential downside of narrowing down what the agent could learn about the system. The instrumental value of detailed microscopic information for a better understanding of the resulting material behavior is known to the material science community, see Refs. \cite{kocks_physics_2003,mecking_kinetics_1981}. In this work, we choose to let the agent have access to that information without any pre-determined and potentially user-biased analysis and interpretation processes like commonly provided in information-condensing characterization protocols.

The second reason for not choosing the classical microstructure characterization measures for the description of the state is that we aim at a domain-ignorant solution, i.e. by an approach that is unbiased by the specific user or specific domain knowledge. Hence, in this work, we consider a more domain-ignorant approach, i.e. we try to find the optimal behavior with as little as possible material science knowledge, i.e. we exploit the full (and sometimes hidden) information content directly from the microstructure images (here from the simulated ones but they could as well be experimentally obtained ones), with much less bias than by classical microstructure characterization.
Along this line, we also use an image to set the goal for the agent, i.e the image of the desired microstructure or the target property. In this way, for example, training and evaluation of the agent can be achieved by measuring the \textit{pixel-by-pixel} similarity between two images, namely, the image of the microstructure obtained by the agent and the image of the target microstructure. Such image analysis, in principle, does not require material-science-specific domain knowledge. Such a domain-ignorant approach increases the range of applicability of our solution to problems where the full domain knowledge does not exist and/or cannot be readily extracted from (modeled or experimentally obtained) images without too much user bias.
 
The last reason for choosing the raw image as a part of the state is that it allows us to find the optimal behavior for reaching the desired microstructure. Being able to target a particular microstructure opens completely new doors to a better and more objective material design, thermomechanical treatment and property development \cite{ashby_materials_2017}.

Using microstructure maps in the form of images as the only state descriptor comes at a potential caveat: obtaining such detailed information might not be always possible during metallurgical processes. Nevertheless, with the rapid progress in the form of advanced in-operando measurement techniques \cite{zhang_situ_2021,zhang_recent_2020,olbinado_recent_2019}, we are getting closer to more continuous coverage of processes through detailed in-situ measurements. One should note that the success of the methodology presented here is not dependent on the particular choice of the state, as long as the state has enough information for the agent to learn the optimal behavior.

The action space, in this work, is designed to consider and reflect those actions which are typically also available in reality. For the specific task chosen here, i.e. microstructure- and energy-sensitive heat treatment, we consequently choose actions such as the increase or decrease of the furnace temperature, maintenance of the temperature at its current value for a specific time, removal of the sample from the furnace (equivalent to shutting the furnace down), etc. Interestingly, such a choice of action space leads to a challenge as there are two different \textit{types} of actions: in particular shutting down the furnace is an action that ends the process, unlike the other actions which are only concerned with regulating the temperature. These types of termination actions are common in optimal stopping problems (or early stopping problems) where the goal is to take an action to stop the problem and otherwise the process continues (see Ref.~\cite{kong2018new} for application of RL to optimal stopping problems). In our case, the difficulty arises from the fact that the continuation action is not one single action but there are three specific choices. In other words, we are dealing with a hierarchical action space, where at the highest level of the hierarchy we need to decide whether to continue or to stop the process. If we decide to continue, then we have to decide at the next hierarchy or decision level, which of the temperature modulation actions we apply. We tackled this problem by the biases induced in the design of the network, as well as the exploration scheme, as explained in the next sections.

The details of the designs of the state space, action space, reward, and termination conditions are listed here:
\begin{itemize}
    \item State space: the state includes the following information: (i) the phase-field as an image of size $(L, L)$ with $L=256$, (ii) the current (scaled) temperature, (iii) the current (normalized) timestep. The scaled variables all fall in the range $[0, 1]$.
    \item Action space: the action space is a discrete space of size four (i.e. consisting of four types of decisions), where there are two types of actions: (i) three actions for temperature regulations, (incremental increase, incremental decrease, or not changing the temperature), and (ii) stopping the process. The temperature, $T$, is in the range $[100, 1000]^\circ$C (approximately pertaining to the typical range of heat treatments of steels) and the allowed increment of the temperature is $\delta T=50^\circ$C.
    \item Reward: as a starting point, we apply a so-called dense rewarding scheme, i.e. giving a non-zero reward at each step. The reward accounts for (i) the similarity between the current phase-field image and the desired phase-field pattern, and (ii) the energy cost at each step. As a quantitative measure for the similarity between the phases, we introduce a new similarity score between the microstructure of the two phase fields (the current one and the desired one) and use the magnitude of its change at each step as a part of the reward. The similarity score is a modified version of the Intersection Over Union approach (IoU). This means that the score has a positive contribution from the overlapping area of the reference phase one from the two images and a negative contribution from non-overlapping areas of the same phase. More precisely, if $C$ and $D$ represent the sets of the current phase 1 and the desired phase 1, the similarity is measured as $|C \cap D| - |(C - D) \cap (D - C)|$~. We normalize this quantity to make learning more reproducible and scaled for the agent. The consumed thermal energy cost per step is calculated by
    \begin{align}\label{energycost}
        e = \alpha(T-T_a),
    \end{align} where $\alpha$ is a constant proportional to the cost of the energy, and $T$ and $T_a$ are the furnace and room temperature, respectively.
    \item The termination conditions are: (i) when the number of steps reaches the maximum number of allowed steps, (ii) when the stopping action is chosen by the agent. 
\end{itemize}

\subsection{Exploration -- \textit{Cautious} Epsilon-Greedy}
As discussed in \secref{env}, we are dealing with a hierarchical action space that is composed of two different types of action, one of them being a process-terminating one. We reflect the nature of this hierarchy in our exploration scheme by modifying the epsilon-greedy exploration plan. In the original epsilon-greedy scheme, to maintain a degree of exploratory behavior, the agent does not act only based on what it has learned but also adds a degree of randomness to its decision-making. In the standard implementation, this randomness gives each of the possible actions the same probability, for example in our case the probability of randomly increasing the temperature is the same as the probability of randomly terminating the process. This implementation leads to undesired exploration features in our case. More specifically it visits the termination action too often and therefore terminates the process before it has a chance of reaching the states in which high similarity would be obtained. Reaching these states, which can happen during long episodes, leads to high rewards and therefore makes the learning of the optimal behavior possible. To avoid this undesired behavior, we introduce a new scheme, which we refer to as \textit{cautious epsilon-greedy}. This new scheme is similar to the established epsilon-greedy approach except that choosing the termination action gets a lower probability compared to the temperature regulation actions. Such short episodes make the training difficult and the proposed scheme is a remedy for that.

Following this rationale in this work, we choose the probability of the termination action to be 20 times less likely than each of the temperature regulation actions. One should note that, although a less frequent choice of the termination action is helping the training in this particular problem, it might be sensible to choose a different probability ratio for other problems. 

\subsection{Neural network design}
The state space and the action space, as explained in \secref{env}, both have specific structures and we reflect these features also in the design of the neural network. In particular,  we first apply several convolutional layers for feature extraction on the image part of the state. Only after that, we combine these (learned) features with the scalar parts of the state using a concatenation operation and further process them using a special fully connected (FC) neural network. This design, which we refer to as \textit{bi-branch} structure, reflects the structure of the action space. The bi-branch structure has two separate fully-connected neural networks which use the aforementioned concatenated features as their inputs. One of the FCs has an output of size one (for the stopping action) and the other has an output of size three (for the three temperature regulation actions). These intermediate outputs are concatenated as the final output of the whole model. 

On the efficiency of the bi-branch structure, we want to emphasize that this is our first attempt to incorporate the structure of this particular hierarchical action space into the late stages of the neural network. More sophisticated designs could be implemented, for example, introducing a hyper-network design~\cite{ha2016hypernetworks} or using a more elaborate combination of the intermediate outputs into the final output (instead of a simple concatenation) would be alternative, refined options. Incorporation of the knowledge about the action space hierarchy into the design of the neural network and the training algorithm is an active field of research in reinforcement learning (for example see Refs.~\cite{xiong2018parametrized, delalleau2019discrete, bester2019multi, hu2021hierarchical}). 

On a practical note, RLlib which is the reinforcement learning library used here for the training of the agent, by default flattens the structure of the state into a one-dimensional array. Nevertheless, the original state space can be restored as we demonstrated in a tutorial (see Ref.~\cite{SIBONI-RLlib-with-Dict-State-2022})

\subsection{Training}\label{sec_training}

For training, we used the RLlib~\cite{liang2018rllib} package as a part of Ray~\cite{moritz2018ray}. The package which is built on top of RLlib together with all the hyper-parameters of the training can be found in the code repository of this work~\cite{SIBONI_RL-Heat-Treatment_2022}. With these parameters, the agent's performance reaches a performance plateau which is an indication of catastrophic forgetting~\cite{mccloskey1989catastrophic, kirkpatrick2017overcoming}. Addressing the reasons and remedies for catastrophic forgetting is beyond the scope of this work, and here we consider the training of the agent to be over right before catastrophic forgetting sets in. The performance of the agent at that stage of the training is presented in the following section.

\section{Results}
\label{sec_res}
In this section, we benchmark the trained agent in controlling the temperature of the furnace to reach the microstructure goal under different conditions. 

\subsection{Microstructure goal}
\begin{figure}[H]
    \centering
    \includegraphics[width=0.9\textwidth]{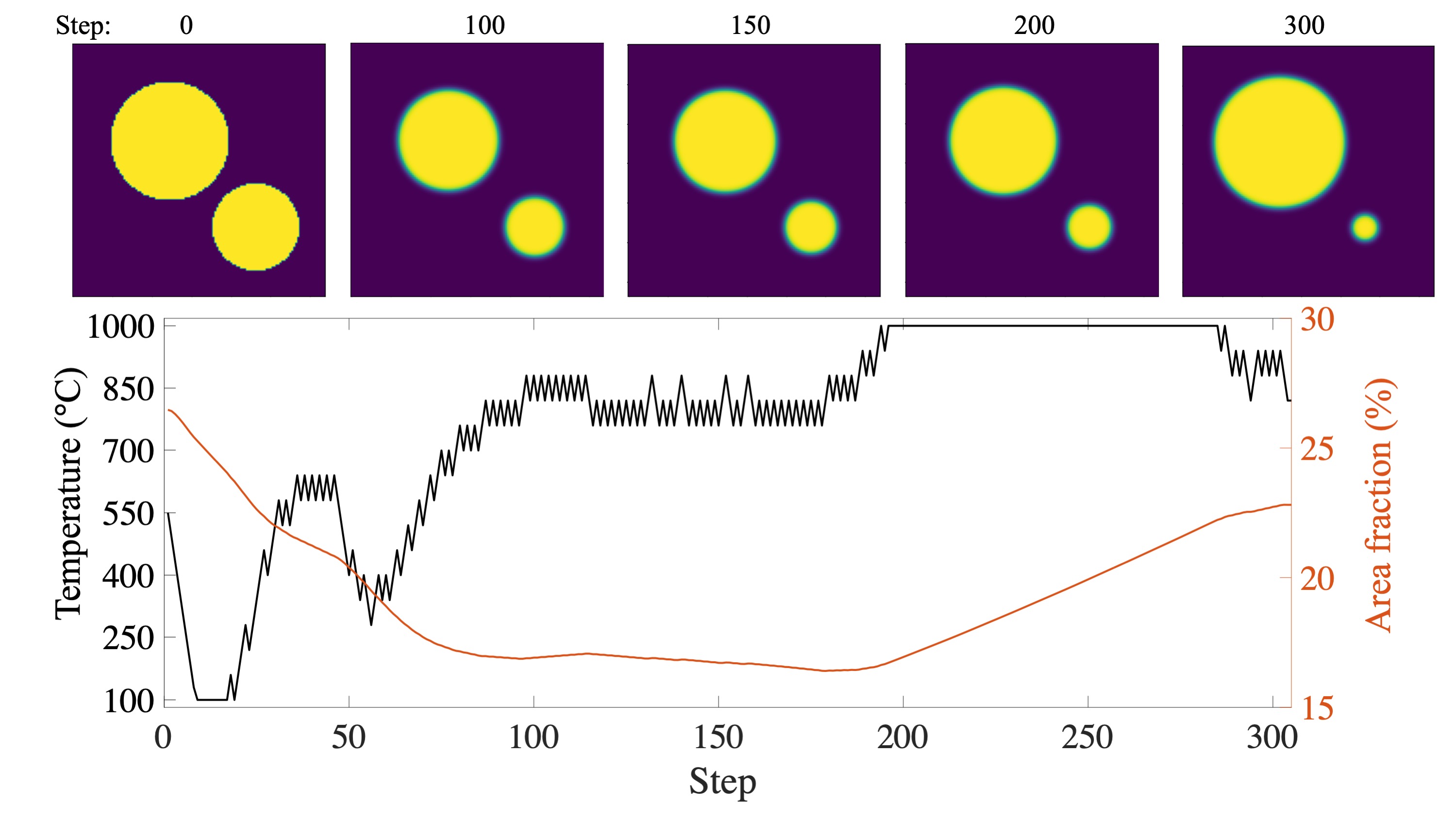}
    \caption{\textbf{An example of the agent's performance in discovering an efficient temperature-time profile to reach the desired final microstructure with minimal total interface length}. Top: snapshots of the microstructure evolution during the heat treatment which is fully controlled by the trained agent. Phases 1 and 2 are shown in dark blue and yellow, respectively. Bottom: The discovered temperature-time profile (black line) and the resulting area fraction (red line) for the system with an initial phase structure of two circular regions. The microstructure goals for the agent are a minimum total interface energy and an area fraction of 20\%. Note that in this particular example, the final area fraction is about 23\%. }
    \label{fig_circle}
\end{figure}

As explained above, the microstructure goals set for the training are here for demonstration purposes the minimum interface energy and the area fraction of 20\% for phase 2. This quantity is a key measure in the phase field model and could translate in real practice for instance to the heat treatment optimization of alloys into the desired microstructure coarsening state. This is for instance an important target in the production of soft magnetic FeSi sheet material used for example in power transformers or in the engine of electrical vehicles.
Once the agent is trained, about 1000 test cases are investigated to benchmark the performance of the agent. A few of these test cases are presented in the following. The first one is an initial microstructure with two circular regions of phase 2 as shown in \figref{fig_circle} in yellow. 

\begin{figure}[H]
    \centering
    \includegraphics[width=0.9\textwidth]{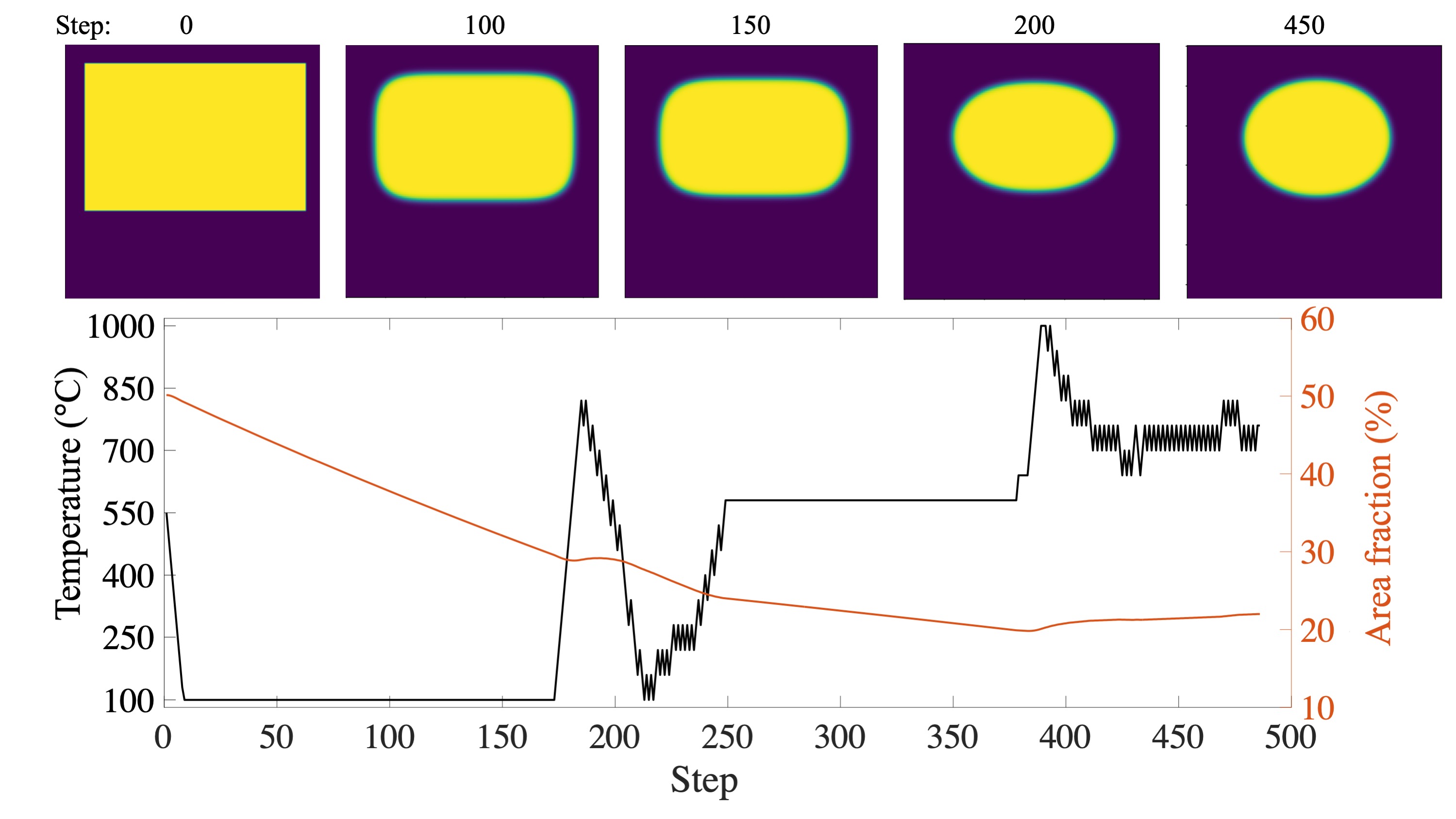}
    \caption{\textbf{An example of the agent's performance in discovering an efficient temperature-time profile to reach the desired final microstructure with minimal total interface length}. Top: snapshots of the microstructure evolution during the heat treatment which is fully controlled by the trained agent. Phases 1 and 2 are shown in dark blue and yellow, respectively. Bottom: The discovered temperature-time profile (black line) and the resulting area fraction (red line) for the system with the initial phase structure of a rectangular region. The microstructure goals for the agent are a minimum total interface energy and an area fraction of 20\%.}
    \label{fig_rec}
\end{figure}

The initial microstructure (step 0 in \figref{fig_circle}) has an area fraction of about 26\% and the interface energy is not at its minimum since it is not a single circular shape. The discovered temperature profile by the agent for this case is shown by the black line in the plot at the bottom of the figure. The resulting area fraction changes are also plotted as a red line together with the microstructure snapshots during the heat treatment. As it is seen, the agent first reduces the temperature dramatically to its minimum value, thus destabilizing phase 2 ($\phi=1$ in \figref{figbulkmob}) in the system. This results in a drop in the area fraction (red line) and shrinking of both circular-shaped regions. The temperature is increased again at around step 20. The subsequent reduction of temperature at around step 50 results in the area fraction reaching the lowest value of about 16\%. At this point, the smaller circular area is no longer stable even at higher temperatures due to its interface energy. The agent decides to increase the temperature and keep it at high values after step 100 to increase the area fraction of the system toward the goal of 20\%. Interestingly, the smaller circular region is still moderately shrinking between steps 150 and 300, while the larger circular region is growing toward the final area fraction goal. This is due to the interface energy and the critical size required for phase stability. To summarize, the agent seems to take advantage of this effect by lowering the temperature dramatically in the beginning, therefore shrinking one of the circles to below its critical stability size. After reaching this point, the agent increases the temperature again to recover the targeted area fraction while the smaller phase region is vanishing. Next, we investigate another special case with a rectangular initial microstructure as shown in \figref{fig_rec}.

\begin{figure}
    \centering
    \includegraphics[width=0.9\textwidth]{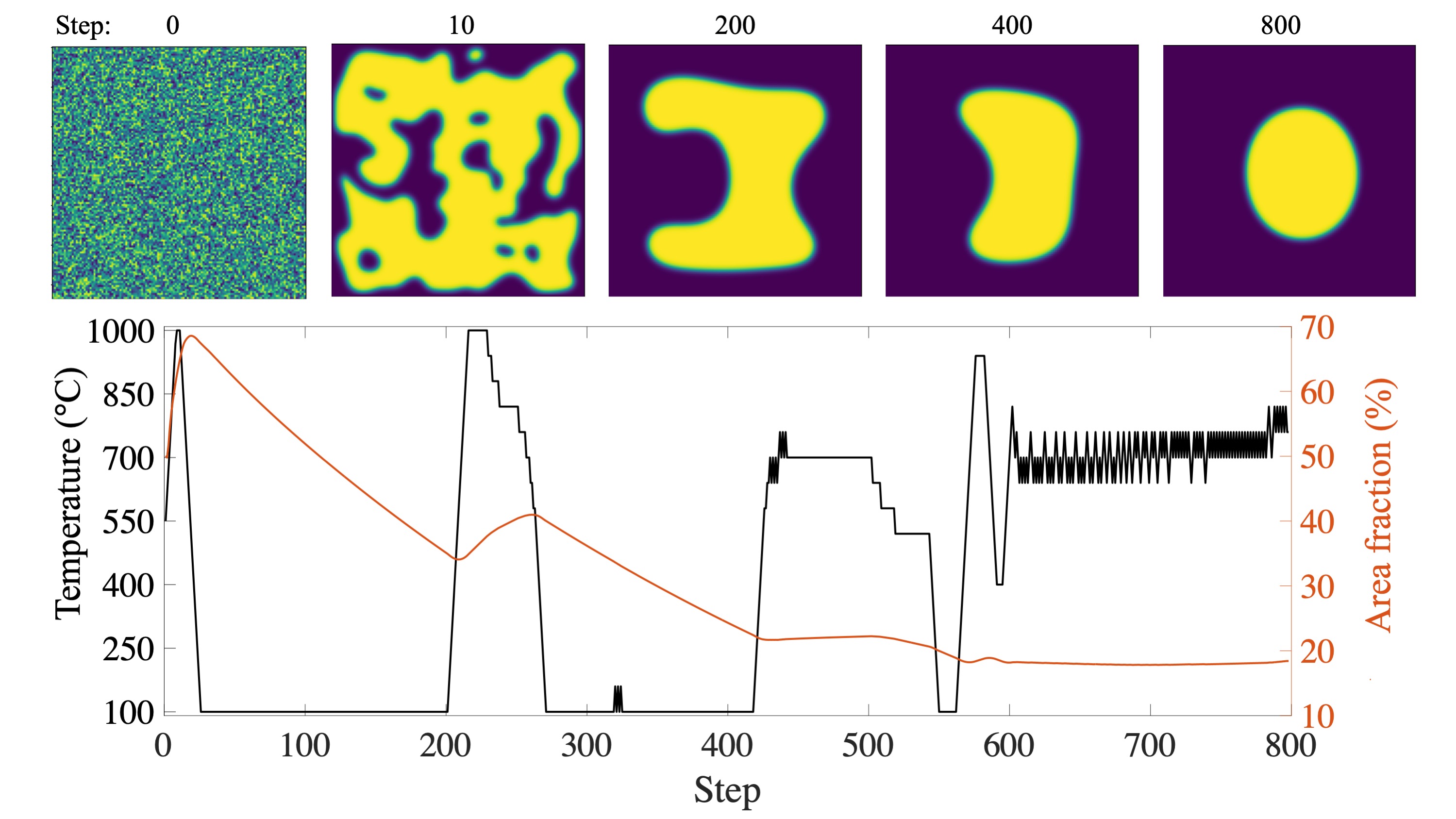}
    \caption{\textbf{An example of the agent's performance in discovering an efficient temperature-time profile to reach the desired final microstructure with minimal total interface length}. Top: snapshots of the microstructure evolution during the heat treatment which is fully controlled by the trained agent. Phases 1 and 2 are shown in dark blue and yellow, respectively. Bottom: The discovered temperature-time profile (black line) and the resulting area fraction (red line) for the system with the initial random distribution of the phase order parameter. The microstructure goals for the agent are a minimum interface energy and an area fraction of 20\%.}
    \label{fig_random}
\end{figure}

Similar to the previously discussed case, here the agent is controlling the temperature profile to reach the same final goals as before. However, the initial microstructure is a rectangular phase as seen in step 0 of \figref{fig_rec} top, with high interface energy and high area fraction. As the initial area fraction (about 50\%) is much larger than the goal of 20\%, the agent immediately reduces the temperature to destabilize phase 2 (yellow). However, as the constitutive material model used in this system is isotropic, the fast reduction of the temperature results in the same boundary velocity in all directions, therefore, the structure essentially maintains its rectangular shape up to step 150. At this point, the agent increases the temperature, thus, slowing down the shrinking process for phase 2. This leads to a longer time for the driving force from the interface energy to move the boundaries toward a more circular shape. From step 250 to about 370, the temperature is kept constant by the agent at about $T = 550 ^\circ$C. This is the temperature at which both phases have almost similar bulk energy (see \figref{figbulkmob}). The agent seems to be able to find this special temperature and exploits the fact that at this temperature both phases are stable. Keeping the system at this specific equilibrium temperature means that there is no driving force from the bulk free energy for driving boundary motion. Therefore, the only driving force will be the capillary-acting interface energy which forces the phase into a circular shape. Toward the end of the heat treatment, the agent increases the temperature again to counteract the shrinkage of the phase, yet, with a small fraction of overshooting, as in the preceding example. The next case investigated in detail from the test dataset is shown in \figref{fig_random}.

As shown in \figref{fig_random}, the initial phase order parameter here is set randomly for all points in the domain with a small fluctuation around 0.5. Therefore, the initial area fraction of the system is 50\%. In this special case, phase 2 (yellow) is dominating at the beginning of the microstructure evolution with its area fraction ramping up to about 70\%. Without the fast reaction by the agent in reducing the temperature at around step 10 to its minimum possible value, phase 2 would have taken over the whole domain. Reducing the temperature also lowers the stability of phase 2 and brings its area fraction down to about 40\% at step 200. This is still significantly above the defined goal of 20\%. Furthermore, the microstructure shape, as seen in the top row of \figref{fig_random} for step 200, is not circular. From this point, the agent increases and subsequently reduces the temperature two times (at around steps 200 to 300, and again from 410 to 550). These changes allow the phase 2 region to change its shape from the concave one at step 200 to a convex and round shape at step 800 which is the defined microstructure goal. 

It should be noted that all cases discussed above are the results of the same agent trained on the same system. It shows that the trained agent can cope with various topological cases and phase distributions. In all the studied cases, it can reach the final goal it had been trained for. Besides the detailed discussion of the agent's decision sequence and action rate for these quite different and specific cases, we benchmarked its performance in about 2000 such cases in the test dataset. These results are discussed in the next section, to go from the understanding of the agent's reaction in a few specific cases to an analysis of its general performance and robustness. 

\subsection{Fluctuation tolerance}
\begin{figure}[H]
    \centering
        \includegraphics[width=0.99\textwidth]{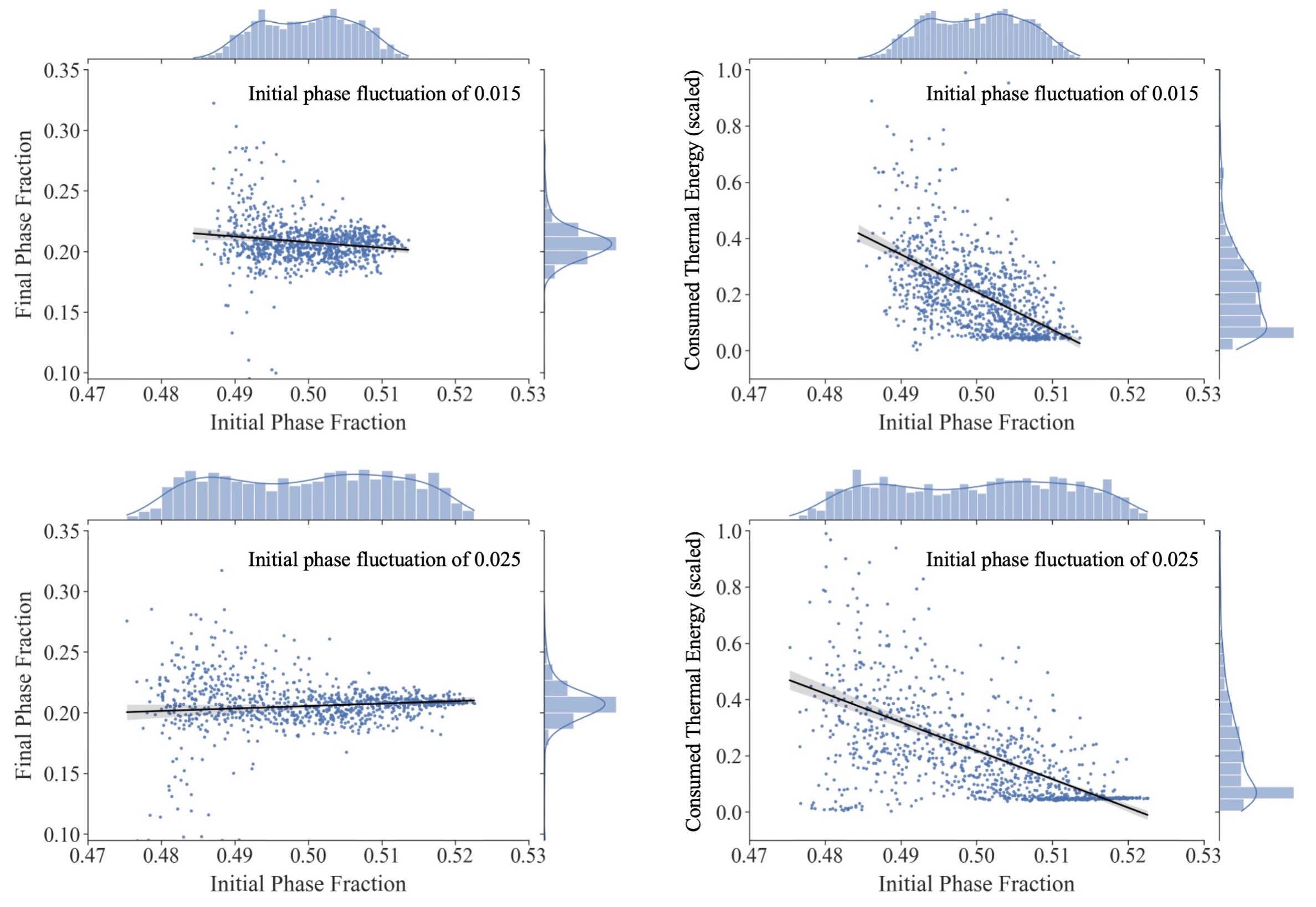} 
        \caption{\textbf{Effect of initial phase fraction fluctuation on the discovery of suited heat treatments by the agent, to reach the target goals of 20\% area phase fraction, the minimization of the interfacial energy, and lowest consumed thermal energy.} Left: the plots show the final phase fraction after application of the heat treatment as a function of the initial phase fraction for the case of lower initial fluctuation (top) and higher initial fluctuation (bottom). The right column corresponds to the consumed thermal energy for each of the test cases.}
    \label{fig_fluc}
\end{figure}

We evaluate the same trained agent discussed above with 2000 test cases where the initial phase distribution is a random field similar to the initial conditions used in training. 
In other words, the order parameter in each point of the simulation domain had been randomly set to a value in the range of $(0.5-\delta\phi, 0.5+\delta\phi)$. The 2000 cases are divided into two datasets, one with $\delta\phi=0.015$ and one with $\delta\phi=0.025$, with the latter having a wider initial fluctuation in the phase values. The results of these tests are shown in \figref{fig_fluc}. The top row is showing the result for the case of $\delta\phi=0.015$ while the bottom row is the result of a larger fluctuation in the initial phase distribution, i.e. $\delta\phi=0.025$. Note that the same agent employed for the examples discussed in detail above is benchmarked here. Therefore, the agent has the same goals of the minimization of the total interface energy in the system and a target area fraction of 20\% with a penalty on the total consumed energy. 

As it is seen in the left column of \figref{fig_fluc}, most of the 2000 test cases successfully reach the final phase fraction goal of 20\%, regardless of the range of the initial fluctuation of the phase fraction. However, as seen in the right column of this figure, the case with a larger initial fluctuation in the area fraction (bottom right) required longer heat treatment and thus more thermal energy input to reach the microstructure goals compared to the case with a smaller initial fluctuation (top right). Furthermore, as it is seen from the fitted black line to the data points, cases with an initial area fraction below 50\% require more energy to reach the microstructure goals as these cases on average require more increase in the temperature. When the area fraction of phase 2 is more than 50\%, phase 2 tends to take over the whole domain. In such cases, the reasonable action is to reduce the temperature of the system, therefore, less energy is used during the heat treatment in such cases as expected. Translating these cases and scenarios studied here into a practical and real-world application case would amount to identifying - with the help of such a trained agent - a heat treatment that is capable of 'repairing' flaws associated with phase inhomogeneity that were inherited from the preceding process steps such as casting and solidification.

\subsection{Energy consumption}
Besides the specific goals of 20\% area fraction and minimum interface energy, used above for demonstration purposes, one of the wider incentives for introducing deep reinforcement learning in the fields of materials processing and manufacturing is to reduce the energy consumption in heat treatment, being one of the main energy sinks in production and manufacturing. As such, in all of the above-discussed cases, the agent has two rewards based on the microstructure (area fraction of 20\% and minimum interface energy) and one penalty on the total consumed thermal energy during the heat treatment. The total amount of consumed energy is simply assumed to be the integral of the temperature-time profile. This will be a good approximation for regimes where most of the energy is spent on the wasted heat of the furnace to the environment, assumed to be linearly dependent on the temperature of the furnace, see \eqref{energycost}. We set the ambient temperature $T_a$ to $22 ^\circ$C and the coefficient $\alpha$ to 0.001 J/$^\circ$C. In this section, we compare the previous network (agent), trained with the above penalty on the total amount of the consumed thermal energy with an agent without this penalty.

\begin{figure}[H]
    \centering
        \includegraphics[width=0.99\textwidth]{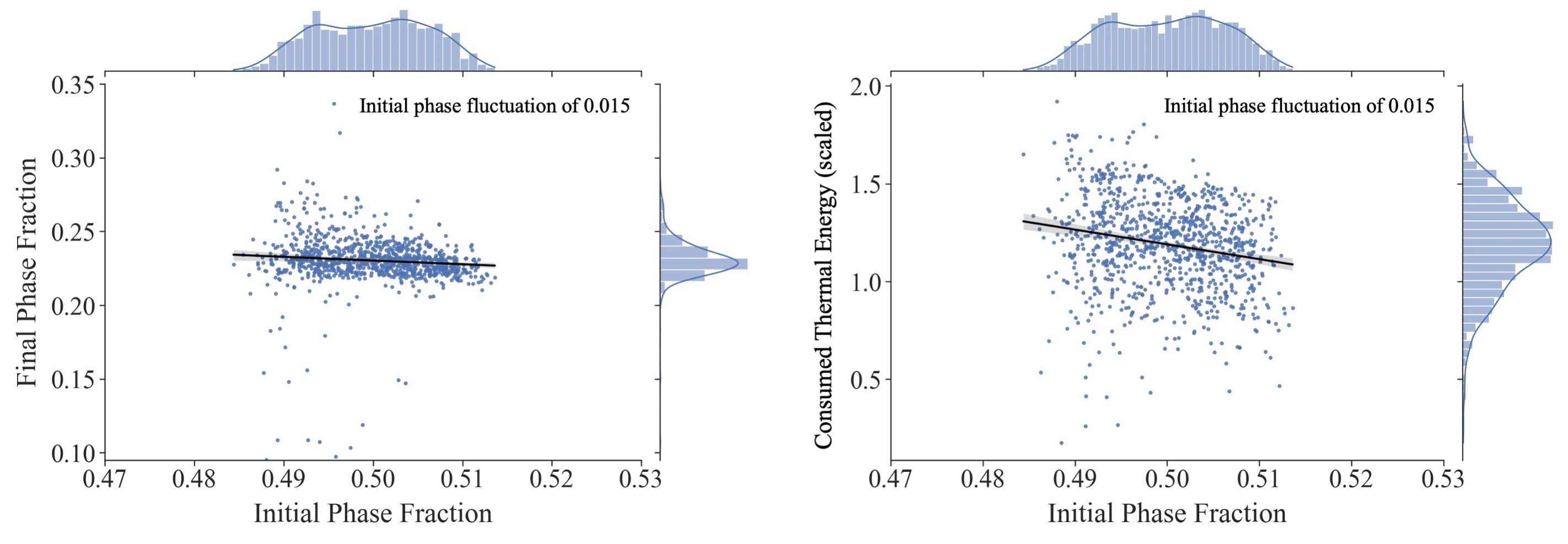} 
        \caption{\textbf{Heat treatment discovery without penalty on energy cost.} Final phase fraction after application of the heat treatment as a function of the initial phase fraction (left). The corresponding energy cost is shown in the right panel. Note that the agent was in this case not explicitly constrained to minimize the total integral of the temperature-time curve.}
    \label{fig_enecon}
\end{figure}

As it is shown in \figref{fig_enecon} left, the microstructure goal is satisfied for various random initial phase fractions. However, the energy cost in this case (right panel in the figure) is significantly (about 2 times) higher than the energy cost in similar tests shown in \figref{fig_fluc}. Since there is no penalty for longer heat treatment profiles, in this case, the agent does not stop the furnace after getting close to the microstructure goal.

\section{Conclusion}
\label{sec_conc}

We showed in this work that deep reinforcement learning (DRL) can be employed in coupling with phase-field (PF) simulations to learn the dynamics and energetics of a material system and control its microstructure. The temperature-dependent PF model is used as an environment for the RL agent to decide on temperature variations, to reach a specific goal, here defined in terms of phase fraction and interfacial energy under the additional constraint of minimum consumed thermal energy. The trained agent is tested in 2000 cases of randomly generated initial states to discover the case-specific temperature profile (heat-treatment profile) to reach the final microstructure target. For a few special cases, we examine the performance of the agent in controlling the system's temperature. We observe that the agent can overcome challenging microstructure geometries through temperature control and in all cases guide the phase structure toward the defined final goal. We also compare two agents, one with and one without a penalty on the total consumed energy. As expected, the agent with a penalty on energy consumption can find temperature profiles suited to reach the microstructure goals at up to 50\% reduced total energy consumption.

\clearpage

\vspace{8mm}
\noindent
\textbf{Data availability}: The data supporting this study’s findings are available from the corresponding author upon reasonable request.
\vspace{4mm}

\noindent
\textbf{Code availability}: 
The implementation of the environment and the reinforcement learning solution (both training and evaluation) in this study are open-source and accessible (see Refs. \cite{SIBONI_Furnace_2022} and \cite{SIBONI_RL-Heat-Treatment_2022}). Other codes and scripts used in this work are available from the corresponding author upon reasonable request.
\vspace{4mm}

\noindent
\textbf{Acknowledgement}: We appreciate the computational resources provided by Max-Planck-Institut f\"ur Eisenforschung GmbH.
\vspace{4mm}

\noindent\textbf{Author contributions}: J.R.M. and N.H.S. developed the initial concept and model formulation. J.R.M. carried out the phase-field model designing and implementation. N.H.S. implemented the DRL using the RLlib package. J.R.M. and N.H.S. wrote the initial draft. D.R. supervised the work and helped in the interpretation of the results. All authors discussed and modified the paper together.
\vspace{5mm}

\noindent\textbf{Competing interests}:
The authors declare no competing financial or non-financial interests.

\bibliographystyle{naturemag}
\bibliography{Refs.bib}
\end{document}